\documentclass{aa}  

\usepackage[]{color}
\definecolor{Red}{rgb}{0.9,0.17,0.31}
\definecolor{Green}{rgb}{0,0.5,0}
\definecolor{ashgrey}{rgb}{0.7, 0.75, 0.71}
\definecolor{bluebell}{rgb}{0.64, 0.64, 0.82}
\definecolor{darkorchid}{rgb}{0.6, 0.2, 0.8}
\definecolor{glaucous}{rgb}{0.38, 0.51, 0.71}
\definecolor{darkcerulean}{rgb}{0.03, 0.27, 0.49}

\usepackage[utf8]{inputenc} 
\usepackage[T1]{fontenc}    
\usepackage{url}            
\usepackage{booktabs}       
\usepackage{amsfonts}       
\usepackage{nicefrac}       
\usepackage{microtype}      
\usepackage{lipsum}
\usepackage[parfill]{parskip}
\usepackage{graphicx}
\usepackage[normalem]{ulem} 
\usepackage{tabularx}
\usepackage[colorlinks, citecolor=darkcerulean, urlcolor=blue]{hyperref}

\usepackage{natbib}

\bibpunct{(}{)}{;}{a}{}{,} 

\title{Encoding large scale cosmological structure with Generative Adversarial Networks}

\author{Marion Ullmo\inst{1,}\inst{2} \and Aurélien Decelle\inst{2,}\inst{3} \and Nabila Aghanim\inst{1}}

\institute{\inst{1} Université Paris-Saclay, CNRS,  Institut d'Astrophysique Spatiale, B\^atiment 121 Campus Paris-Sud 91405, Orsay, France \\
\inst{2} Université Paris-Saclay, CNRS, TAU team INRIA Saclay, Laboratoire de recherche en informatique,
91190, Gif-sur-Yvette, France. \\
\inst{3} Departamento de Física Téorica I, Universidad Complutense, 28040 Madrid, Spain
}

\authorrunning{A1}
\date{}

\abstract{ Recently a type of neural networks called Generative Adversarial Networks (GANs) has been proposed as a solution for fast generation of simulation-like datasets, in an attempt to bypass heavy computations and expensive cosmological simulations to run in terms of time and computing power. In the present work, we build and train a GAN to look further into the strengths and limitations of such an approach. We then propose a novel method in which we make use of a trained GAN to construct a simple autoencoder (AE) as a first step towards building a predictive model. Both the GAN and AE are trained on images issued from two types of N-body simulations, namely 2D and 3D simulations.
We find that the GAN successfully generates new images that are statistically consistent with the images it was trained on. 
We then show that the AE manages to efficiently extract information from simulation images, satisfyingly inferring the latent encoding of the GAN to generate an image with similar large scale structures.}

\keywords{(Astronomical instrumentation, methods and techniques:) Methods: statistical, Methods: data analysis}

\begin{document}


\maketitle

\section{Introduction} \label{Introduction}

The standard cosmological model provides a description of the Universe as a whole: its content, its evolution and its dynamics. In this model, the structures observed today (galaxies and clusters of galaxies) have evolved from tiny fluctuations of density imprinted during the very early stages of the Universe \citep{LSS1,LSS2,LSS3}. Matter has over time accreted from this homogeneous distribution to form what is today a complex network of structures known as the cosmic web \citep{bond1996filaments}. This hierarchical assembly of matter was tested thanks to increasingly large numerical simulations such as Millennium \citep{Millenium} and Illustris \citep{Illustris}, and confirmed with actual observations of large-scale matter distribution in galaxy surveys such as the Sloan Digital Sky Survey (SDSS) \citep{SDSS}. 


However, the large and detailed simulations including detailed baryonic physics \citep{dubois2016horizon,mccarthy2016bahamas} that are needed to compare theory with observations are computationally expensive. Faster fully analytical approaches \citep{shandarin1989large,kitaura2013cosmological} and semi-analytical simulations \citep{monaco2002pinocchio,tassev2013solving} relying on first or second-order perturbation theory exist, but they cannot address the highly non-linear stages of the structure formation. 

The recent advances in computer technology and in machine learning have prompted an increasing interest from the astronomical community in proposing machine learning as an interesting alternative for fast generation of mock-simulations and mock data, or for image processing. Indeed, the ever larger quantities and quality of astronomical data calls for systematic approaches to properly interpret and extract the information which can be based on machine learning techniques such as in \cite{villaescusa2020camels},\cite{schawinski2018exploring}, and \cite{bonjean2020deep}. Machine learning can also be used to produce, in a computationally cheaper manner, density maps from large N-body simulations of the dark matter \citep{rodriguez2018fast,feder2020nonlinear} , or to infer a mapping between the N-body and the hydrodynamical simulations without resorting to the full simulations \citep{troster2019painting,zamudio2019higan}.

In this context, certain types of neural networks called \textit{convolutional neural networks} (CNN)\citep{lecun1990handwritten} excel in the general field of image processing thanks to their automatic pattern detection property (for a comprehensive review, see \cite{ntampaka2019role}). 
Among the CNNs, the generative models such as the Generative Adversarial Networks (or GANs) \citep{goodfellow2014generative} have shown promising results both in computer science and physics \citep{casert2020optical,de2017learning,ahdida2019fast}. These networks aim to learn a probability distribution as close as possible to a considered dataset's in order to later generate new instances that follow the same statistics. GANs have proven to be very promising tools in terms of media generation \citep{GANsound,GANvideo}. In Astronomy, they have recently been used in several cases and more specifically by \cite{rodriguez2018fast} and \cite{feder2020nonlinear} to provide a fast and light alternative to simulations and images.

In the present work, we build on the studies of \cite{rodriguez2018fast} and \cite{feder2020nonlinear} and explore the use GANs to generate 2D and 3D-based images of N-body simulations. Furthermore, we present how we can make use of a trained GAN to construct an \textit{Autoencoder} \citep{hinton2006reducing}. This network relies on the basic concept of intaking data of a certain dimension (typically an image in the form of an $n\times n$ matrix), learning a representation of reduced size of the data (typically a vector of size $m<<n\times n$), and providing a key to decode this representation. In Sec. \ref{sec:data}, we describe the data used to train the neutral networks, from the simulations they are extracted from to the way in which we construct them. In Sec. \ref{networks}, we present in detail the two types of networks that we will use for our work, then in Sec. \ref{results} we present the results we obtained on both network types for both types of image. Finally we will conclude on our findings.

\section{Data}\label{sec:data}

Our work rests on the use of 2D convolutional networks which are applied on \textit{images}, which are simply two-dimensional arrays, with every element representing a \textit{pixel} and its value representing the pixel's intensity.

Given that we are using 2D convolutional networks, we want to test them on compatible 2D images; therefore we will be conducting tests on both images built from 2D simulations for a "best case scenario" as well as images from slices of 3D simulations, which are of interest to us for further work. Both sets will \textit{in fine} be two-dimensional images since the 3D dataset will be sliced and stacked on small redshift, $z$, intervals. For simplicity we will refer to them as 2D images and 3D images.

\begin{figure*}
  \centering
  \includegraphics[width=.8\textwidth]{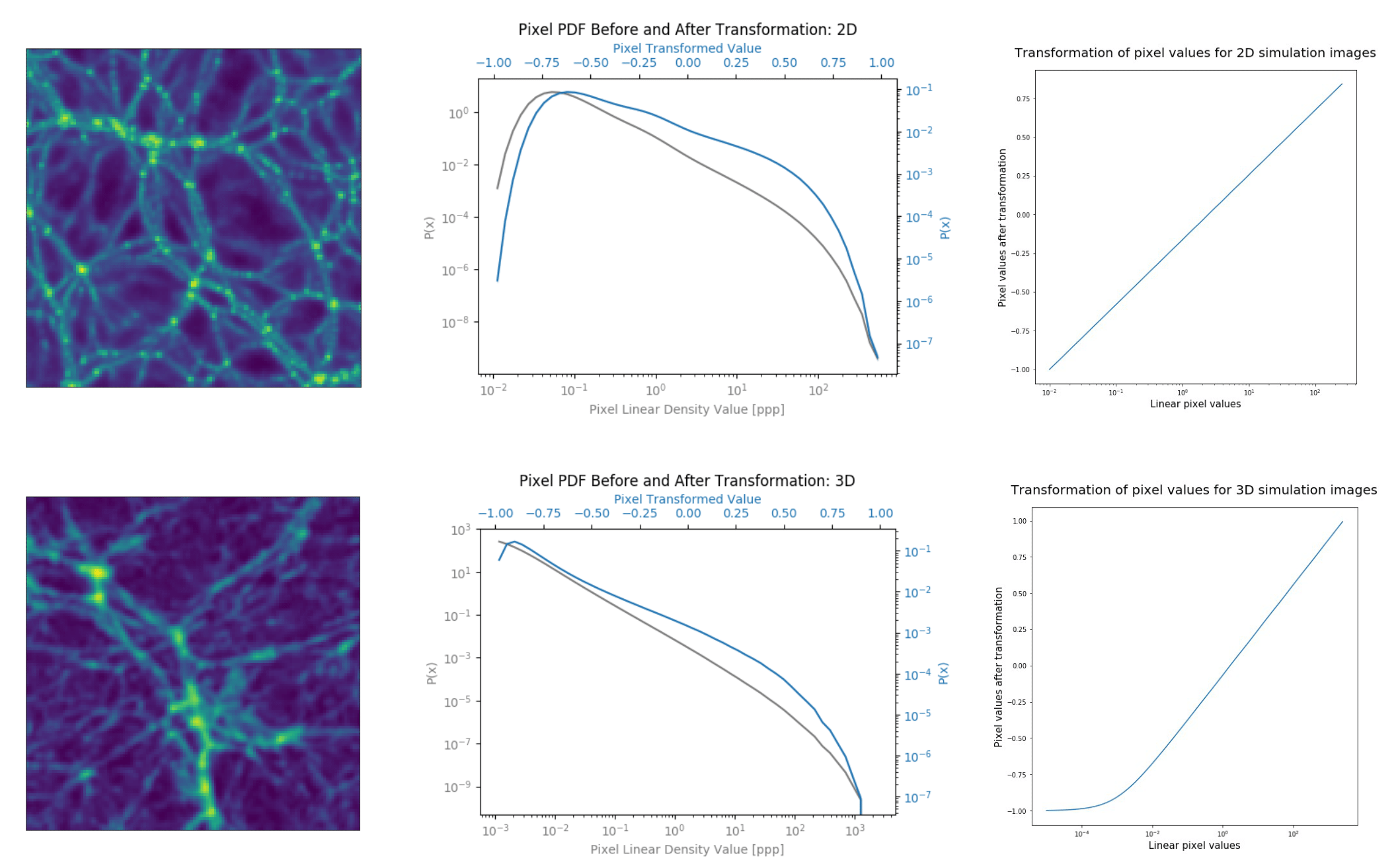} 
  \caption{ Example of a simulation image (left), histogram of the pixel values before and after log-like transformation (middle), and pixel value transformation function (right) for the 2D (top) and 3D (bottom) images. In the middle column, grey represents the pixel values histogram in linear scale and blue the pixel value histogram after log-like transformation (top: eq. \ref{eq:marion_tf} and bottom: eq.\ref{eq:rodriguez_tf}) of the images. For both cases the GAN has been trained using the log-transformed sets of images.}
  \label{fig:pixelpdf}
  
\end{figure*}

\subsection{Simulations} \label{sims}


The 2D images are produced from a publicly available 2D particle-mesh N-body simulation code \url{https://zenodo.org/record/4158731#.X5_ITJwo-Ch} to simulate 1000 "boxes" of size $(100 Mpc/h)^2$ with $512^2$ particles using the standard $\Lambda CDM$ cosmology. 

The 3D data used for this analysis are snapshots from numerical simulations of large scale structures produced with the publicly available code GADGET2 \citep{GADGET1,GADGET2}. These are dark matter (DM) only simulations, referred to as N-body simulations. GADGET2 follows the evolution of a self-gravitating collisionless particles. This is a good description of DM dynamics in accordance with the cosmological model, since DM only interacts gravitationally with itself as well as with baryons. In practice, the GADGET2 code computes gravitational forces with a hierarchical tree algorithm to reduce computing time and to avoid having to compute the gravitational effect of $N$ particles on each particle (which would mean $N^2$ computations at each time step). The algorithm divides space using a grid. Then to compute the gravitational forces exerted on an individual particle, GADGET2 groups particles more and more coarsely according to their distance and computes the gravitational pull of the groups rather than that of the individual particles.


The simulation starts at redshift $z=99$ with a 3D box of 100 Mpc$^3$ size (chosen to contain representative large-scale structures) with a quasi homogeneous distribution in space of $512^3$ DM particles, with Gaussian distributed very low-amplitude inhomogeneities, and an initial velocity associated with each particle. The inhomogeneities stand for the initial density perturbations produced in the early Universe that will eventually evolve into galaxies, clusters and filaments. The system is then evolved with the particles only being subject to gravity. Cosmic expansion is also taken into account and we use the cosmological parameters $\Omega_\mathrm{m}$: $0.31$, $\Omega_\Lambda$: $0.69$, and $H_0$: $0.68$ from Planck 2018 \citep{aghanim2018planck}.
The simulation is run up to the the present epoch ($z=0$). At any time step, we can retrieve the individual particles' positions and velocities in the 3D box. These data, describing dynamical state of the system at a particular time, are referred to as a \textit{snapshot}. To build our dataset, we only retain the positions. They will be used as inputs for the network.

\subsection{Construction of the sample} \label{sample construction}

We first use the set of 1000 native 2D simulations. 
From these 1000 independent discrete $256\times256$ density maps are obtained by estimating local densities from 2D snapshots with the help of a Delaunay tessellation field estimator \citep{aragoncalvo2020smooth}. We use this as a basis to construct the images.

We aim to build a similar set of 2D density maps from our 3D ($z=0$) snapshot. To do so, we define along a given axis slices of size $\Delta z = 0.3$ Mpc. We hence define $333$ slices, and repeat the process for the other two axes and obtain a total of $1000$ slices. 

Every slice is then converted into a 2D discrete density map by computing the histogram of particles over a $770\times770$ grid. 
After applying a log-like transformation (see Eq. \ref{eq:rodriguez_tf}), the grid is smoothed with a Gaussian filter of standard deviation the size of three pixels with a stride of three pixels. This choice yields images in which structures are smooth while preserving the fine low-density structures and hence results in significantly better results than standard stride-less Gaussian smoothing when used with the different networks in our study. Eventually, we construct a set of 1000 $256\times256$ images that will be used in the following analysis.

For both sets of 2D and 3D-based images, we also proceed with data augmentation in order to increase the number of images, and hence the size of the training set of our network. By rotating and flipping each image we multiply the size of our dataset by eight. We also choose to use $128\times128$ images extracted from the $256\times256$ ones. These images of side length $50 Mpc$ capture the largest cosmic structures. In doing so we produce nine smaller images from a single initial one. Eventually, we construct a dataset of 72000 $128\times128$ images which we will use as a basis for the training.


The GANs operate by using a set of filters (Sec. \ref{GAN}) to recognize and learn patterns at different size scales in an image. Therefore we need to work with images with clearly apparent patterns such that the GAN can easily detect the set of images' salient features. However linear density maps of the cosmic web show a poor array of shapes, with images appearing mostly uniformly dark with occasional bright pixels corresponding to dense halo centers; on the other hand a log representation of the same density maps make the cosmic web's filaments apparent, providing shapes and texture that the GAN can more readily detect and reproduce.
Furthermore GANs are built to intake, and create, images with pixel values $\in [-1,1]$.


We hence apply a log-like transformation to the images' pixel values (\ref{fig:pixelpdf}).


For the 2D images, the pixel value $v'$ in the "transformed" images writes:

\begin{equation}
v'=\frac{2\log(v)-(b+a)}{b-a} \label{eq:marion_tf}
\end{equation}

$v$ is the original pixel value, $a$ and $b$ are chosen such that $a \lesssim \min(\log(v))$ and $b \gtrsim \max(\log(v))$ so as to have $v'\in ]-1,1[$ compatible with the network inputs. We use these strict inequalities to give the network freedom to exceed its training set's boundaries when generating images with pixel values $\in [-1,1]$. We thus set $a=\log(0.01)$ and $b=\log(600)$.


The images from the 3D simulations have a significantly larger range than the 2D ones with values up to $2000$ particles per pixel as well as zero values.
For these images, we adapt the above-described transformation to better suit the images' pixel range and 0 values. We recall that this transformation is applied \textit{before} the smoothing with a Gaussian filter. The obtained pixel value $v'$ writes:

\begin{equation}
v'=\frac{2\log(v+c)-(b_{2}+a_{2})}{b_{2}-a_{2}} \label{eq:rodriguez_tf}
\end{equation}

where $v$ is the original pixel value, $b_{2}$ is chosen such that $b \gtrsim \max(\log(v))$, $c$ is chosen such that $c> 0$ but $c<<\bar{v}$ to increase the contrast, while allowing for a log transform, and $a_{2}= \log(c)$. We set $b_2=\log(2632)$, $c=0.001$ and $a_2=\log(0.001)$.

It worth noting that while adding the constant $c$ allows for a log-like transformation, it makes the linear values smaller than $c$ difficult to distinguish from one another after transformation. This can be observed in the lower right panel of Fig. \ref{fig:pixelpdf} which represents the transformation function given by Eq. \ref{eq:rodriguez_tf}. We clearly see a saturation effect for values below $c=10^{-3}$. We therefore do not expect our networks to recover the pixel pdf correctly for values below $c$.



\section{Unsupervised Neural Networks} \label{networks}

In the following, we will use two types of neural networks, described in more detail below. First, we will train a Generative Adversarial Network (GAN) as a means to recover our images' underlying distribution and generate new images hailing from this distribution.

Second, having learned a representation of our image sets with the GAN, we will build upon it to create an Autoencoder that can recover the latent encoding of any given image within this representation. 

Concretely, this will create a tool which can extract any image's essential information in the form of a vector of small size (100) and recover the same image with this information. This property to extract key information can then further be used for more complex purposes, such as prediction or detection.



Both networks are more specifically convolutional neural networks, or CNNs. A neural network can be described as a differentiable function $N$ with parameters (or weights) $\theta_N$, input(s) $x$ and ouput(s) $N(x)$. As with other machine learning models, it is trained to complete a specific task by minimizing a loss function $L_N(\theta_N,x)$ by gradually modifying the parameters to reduce the loss over multiple iterations with different $x$.

A CNN builds or extracts information from images through a series of convolutions between these images and small filters which act as feature detectors. Training of the network rests on learning a set of filters that optimally detect or recover a set's defining shapes and structures.


 
\subsection{GAN} \label{GAN}

Given a dataset to train on, GANs extract the underlying modes of its distribution and can then generate new data sharing the same distribution and thereby similar to the training dataset. Trained correctly, GANs can hence be used to produce an infinite amount of new images given a large but finite number of input images (i.e. training dataset). 

The GAN consists of two competing neural networks. The first, a \textit{Generator} produces images taken as input a random vector. The second network, a \textit{Discriminator}, tells apart these generated images from \textit{true} ones from the training set. As both networks start out with no information about the images, the tasks of both Generator and Discriminator start out as simple: the Generator easily "fooling" the Discriminator and the Discriminator having to tell apart very dissimilar images. However as each of the two networks becomes more efficient, one at generating convincing images and the other at differentiating them from the true set, the task is made harder for the other network. Through this competition both networks train each other by gradually increasing the difficulty of the other's task while simultaneously improving themselves.


The networks work in the following way. The Generator takes a random Gaussian-distributed 100-vector ($z$) as input and from it builds an image ($G(z)$) through a series of deconvolutions and activations as described in Appendix \ref{sec:appenA}.
The Discriminator takes in an image, either from the training set ($x$), or from the set produced by the Generator ($G(z)$), and through a series of convolutions and activations further described in Appendix \ref{sec:appenA} yields a single number $D(x/G(z)) \in [0;1]$ which represents the probability with which it estimates the input image to come from the training set.

The training procedure can therefore be described as follows. First, the Generator will generate a batch of images (in our case 50) and the same amount of images will be drawn from the training set. Then, the parameters $\theta_D$ of the Discriminator will be adjusted such that, the probability given by the Discriminator that the generated images are the \textit{true ones} decreases, while at the same time the same probability computed on the training set increases. Denoting $m=0,\dots,50$ the indices of the generated image $z^{(m)}$ and the training set images $x^{(m)}$ in the batch, we want to maximize the following loss

\begin{equation}\label{eq:loss}
    l_G = \prod_m D(x^{(m)}) \prod_m (1-D(z^{(m)}))
\end{equation}

which is equivalent to minimizing the following log-loss
\begin{equation}\label{eq:logloss}
    L_D=-\frac{1}{2}\mathbb{E}_x \log D(x)-\frac{1}{2}\mathbb{E}_z \log(1-D(G(z)))
\end{equation}

where $\mathbb{E}_x$ represent the average over the dataset and $\mathbb{E}_z$ the average over the random vector $z$. To minimize this expression, the Discriminator should yield a prediction near to one for the images of the training dataset and near to zero for the images produced by the Generator. Conversely, the Generator should aim at producing images that look like "true" images, and so for the Discriminator to yield predictions close to one when assessing its generated images. Therefore the Generator's loss is simply defined as:

\begin{equation}
    L_G=-L_D
\end{equation}






At the end of the training stage, the two networks should converge to an equilibrium wherein the Discriminator is unable to distinguish between the two sets of images and the Generator is outputting images sampled from the training set's true underlying distribution.

In practice, most GANs, including ours, never perfectly reach this equilibrium and instead reach a point where the quality and diversity of the generated images fluctuates with training. Therefore instead of stopping training and collecting the resulting networks at a specific point we elect to save the weights of our networks regularly during training and choose the best set of weights by comparing the quality of images they generated and their statistical properties.

This done, we are equipped with a functioning Generator and Discriminator which we will further use in the construction and training of an Autoencoder.


\subsection{Autoencoder} \label{AE}

An autoencoder (AE) is a neural network that learns in an unsupervised manner a representation (encoding) for a set of data, typically for dimensionality reduction \citep{hinton2006reducing} or latent encoding \citep{makhzani2015adversarial}.


A basic autoencoder is built and trained in the following way. A first network $e$, called {\it Encoder} takes in an input $x$ and outputs a vector of reduced size $z=e(x)$. A second network $d$, the {\it Decoder}, takes input $z$ and outputs $\tilde{x}=d(e(x))$. Now the AE is trained by imposing that the resulting $\tilde{x}$ is as close as possible to the initial $x$. This is typically (but not systematically) done by using a cross-entropy loss, or making $\| x - d(e(x)) \|^2$ to be as small as possible.

Incidentally, we can also consider and use the GAN's Generator as a readily trained decoder from a reduced space ($R^{100}$) to our target space ($z=0$ simulation slices). Indeed, the Generator has learned a representation of the simulated images. Taking in an input $z$ of reduced size $100$, the Generator is be able to output any image $\tilde{x}=g(x)$ from the simulations. Furthermore, given the Generator's ability to generate images that are statistically consistent with those of the simulations, using it in the AE would constrain the outputs to share the same statistical properties. Thus we would avoid the AEs' common pitfall to output blurry images  \citep{dosovitskiy2016generating}.

To create a functioning AE, we therefore need only build and train a functioning Encoder that will work as an inverse function of the Generator, such that $\tilde{x}=g(e(x))=g(g^{-1}(x))=x$. 

It bears mentioning that in a classical autoencoder, the encoder and decoder are trained alongside each other. In our scenario the decoder is trained first separately in an effort to constrain it to output images that are statistically sound (ie that hail from the simulation images' underlying distribution). These constraints might imply a loss of accuracy in our recovery of structures in individual images. It is therefore capital to look at both global statistics and pairwise to see how the autoencoder fares in both regards, as is done in Sec. \ref{results}. 




Our Encoder is built as a simple convolutional network which reduces an image to a vector of size $100$ in order to match the Generator's input. The different layers of the Encoder are based on the Discriminator's architecture, since the latter is especially developed to extract essential information from simulated images. However since the goal of the network differs from the Discriminator's we will only retain the architecture and not the weights. Further details on the architectures of the networks can be found in the Appendix \ref{sec:appenA}. 


\begin{figure*}
  \centering
  \includegraphics[width=\textwidth]{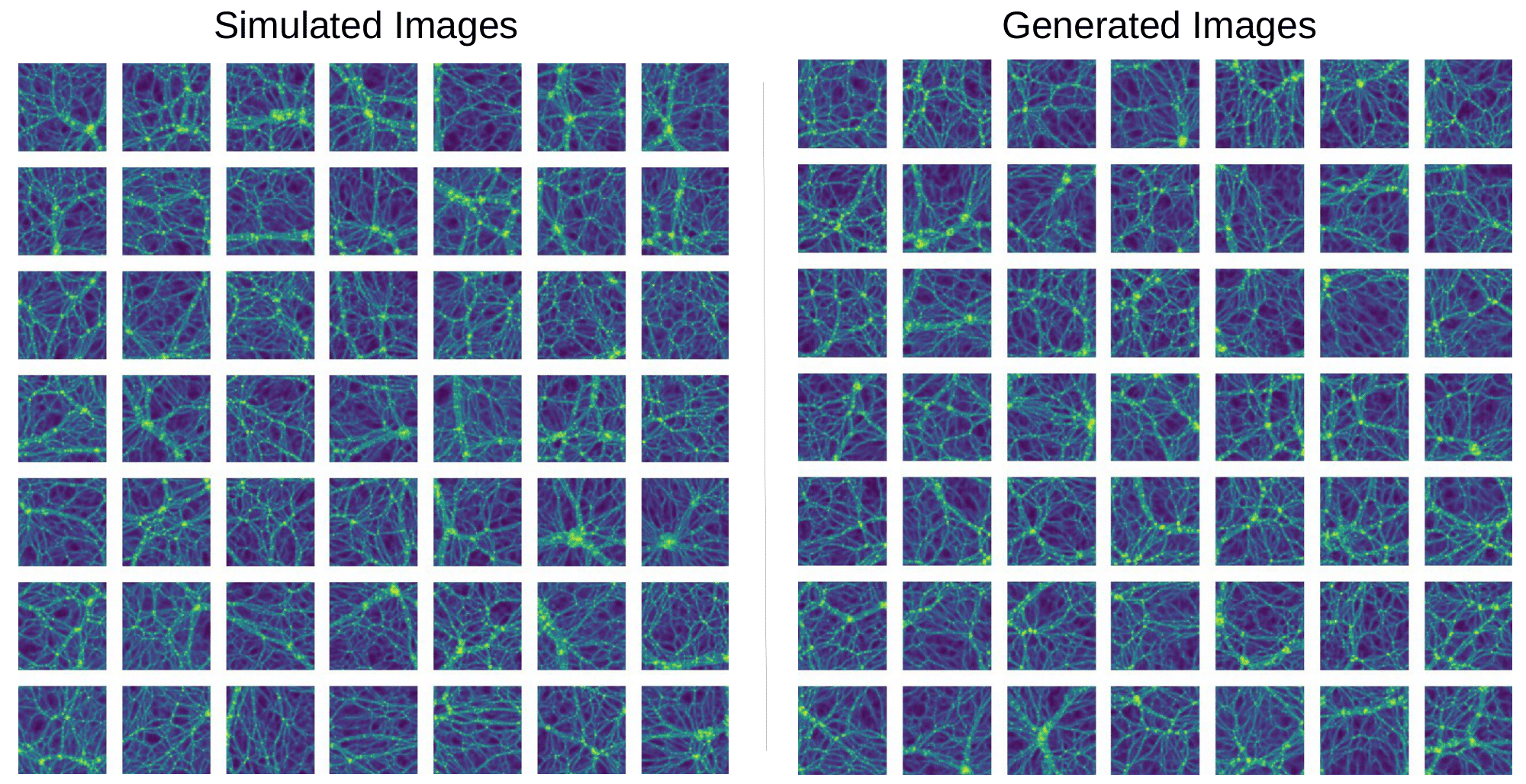} 
  \caption{Two subsets of 50 images taken at random from a set of 2D simulation images (left) and a set of images generated by the GAN (right). Every image represents a $128 \times 128$ log density map of side $50 Mpc$. They are virtually indistinguishable by eye.}
  \label{fig:2D_GAN_ims}
  
\end{figure*}

We can now build the AE by putting the two networks (Encoder and Decoder) end-to end. This can be described as the following function: $a(x)=d(e(x))$. We fix the weights of the Decoder and update the weights of the Encoder to decrease the loss function previously described:  $\| x - a(x) \|^2$

However instead of using the $\ell_2$ loss in the image's space, comparing pixels at the same location on both images, which accounts poorly for well-recovered but slightly shifted structures, we instead make use of the our Discriminator. It is expected that the Discriminator manages to learn a latent representation of our dataset's images during the GAN's training. This latent representation is given by the penultimate layer, where its elements are used to estimate the probability to be or not a \textit{true image}. This representation of an image in the Discriminator's latent space is semantically meaningful \citep{bang2020discriminator}, accounting for the presence of specific structures or shapes, and tends to put visually similar images at a small distance in this space, where they would otherwise be more distant in the images' space.

Therefore we define our autoencoder's loss on an image as:

\begin{equation}
    L_{AE}=\Delta(x,\tilde{x})
\end{equation}

where $\Delta$ is the $\ell_2$ difference in the discriminator's latent space. Or, calling $TD$ the truncated discriminator with its final layer removed:

\begin{equation}\label{eq:TDloss}
    L_{AE}=\| TD(x) - TD(\tilde{x})) \|^2
\end{equation}

We can then train the AE by updating its weights to minimize this loss. Training is stopped when the loss measured on a separate {\it validation} set reaches a minimum.

From now on we will refer to the images $\tilde{x}$ reconstructed by the autoencoder as \textit{inferred images}. We will assess their quality in Sec. \ref{results}.

\begin{figure*}
  \centering
   \includegraphics[width=\textwidth]{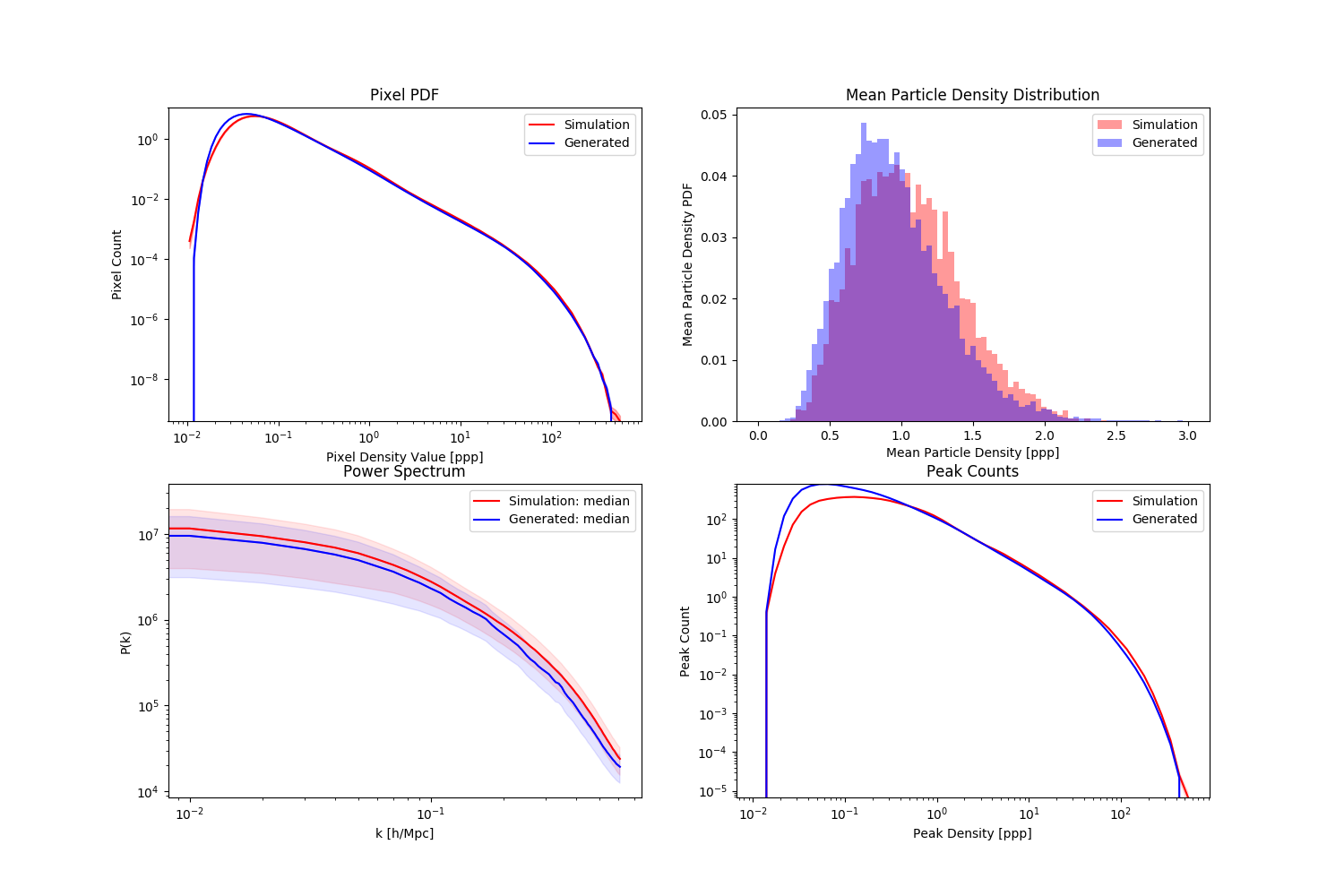} 
  \caption{Statistics of the 2D simulation images compared to their GAN-generated counterparts. \textbf{Upper left} shows the pixel PDF, \textbf{upper right} shows mean density distribution, \textbf{lower left} shows median power spectrum as well as the median  absolute deviation (mad) layer, and \textbf{lower right} shows average peak count per image.}
  \label{fig:2D_GAN_stats}
  
\end{figure*}

\section{Results} \label{results}

In this section,  we use a set of statistical estimators to evaluate the performances of both the GAN and the autoencoder in respectively emulating or reproducing the input images  from 2D or 3D simulations, considered here as the "ground truth". Given that we ultimately want to reproduce the ground truth linear density maps, we apply an inverse-log transformation to revert to linear density. This density, corresponding to the pixel values, will be expressed in \textit{particles per pixel} (hereafter noted ppp) and referred to as $\rho$.

We compare the sets of images on the basis of their pixel PDFs, the distribution of their mean density, their power spectra, and their peak counts.
Additionally, as we expect the AE's inferred images to reproduce their input images, we perform a pairwise measure described in Sec. \ref{overlap} to quantify how well the individual images are recovered.



\subsection{Statistical estimators} \label{Estimators}
\subsubsection{Pixel PDF and Distribution of the mean density} \label{PDF}

A first basic test is to compare the distributions of pixel values, which correspond to a density measure in particles per pixel (ppp) in both sets of images. We also compute the mean particle density of each image $\mu$ and compare their PDF over the simulated (truth) sets and generated and inferred image sets from the GAN and AE.

Whereas the pixel PDF is informative of the density distribution of an image on average, and therefore ensures that two sets of images are similar on average, the mean density serves as a simple one-dimensional visualisation of the distribution of images over a set. This type of information is important to ensure that we recover both datasets' underlying distribution, and recover different cosmic regions and different halo densities in the right proportions.


Furthermore GANs can often suffer from "mode collapse" \citep{thanh2018catastrophic}, a situation where the images generated are indistinguishable from the original set but show little to no diversity. 

Although visual inspection of the images can help to exhibit mode collapse, a visualisation of the overall distribution through the mean density provides additional information to confirm its absence.







\subsubsection{Peak counts} \label{peakcounts}

We compute the average peak counts over the generated or simulated dataset. A peak is a local maximum ($\rho_{
\rm max}$) defined as a pixel whose eight contiguous neighbors are of smaller values. For each image, we compute the number of peaks for a given value of $\rho_{\rm max}$, and average this number over the whole dataset.

In the simulated images, the higher peaks, being dense local maxima, are expected to correspond with halo centers, whereas smaller near-zero peaks are more likely to be the result of noise from the simulation or the image-making process. Therefore, we are more interested in the former, which give us an indication as to our recovery of halo distribution.

\subsubsection{Power Spectrum} \label{PS}

We compute the power spectrum of each image from the sets (input, generated and inferred). For a frequency $\nu$ it is given by:

\begin{equation}
P(\nu)=\langle \|A_{kl}\|^2 \rangle_{(k,l) | k^2+l^2=\nu^2}, 
\end{equation}

where $A_{k,l}$ are the image's discrete Fourier transform elements.

\subsubsection{Sørensen–Dice coefficient} \label{overlap}

\begin{figure*}
  \centering
  \includegraphics[width=\textwidth]{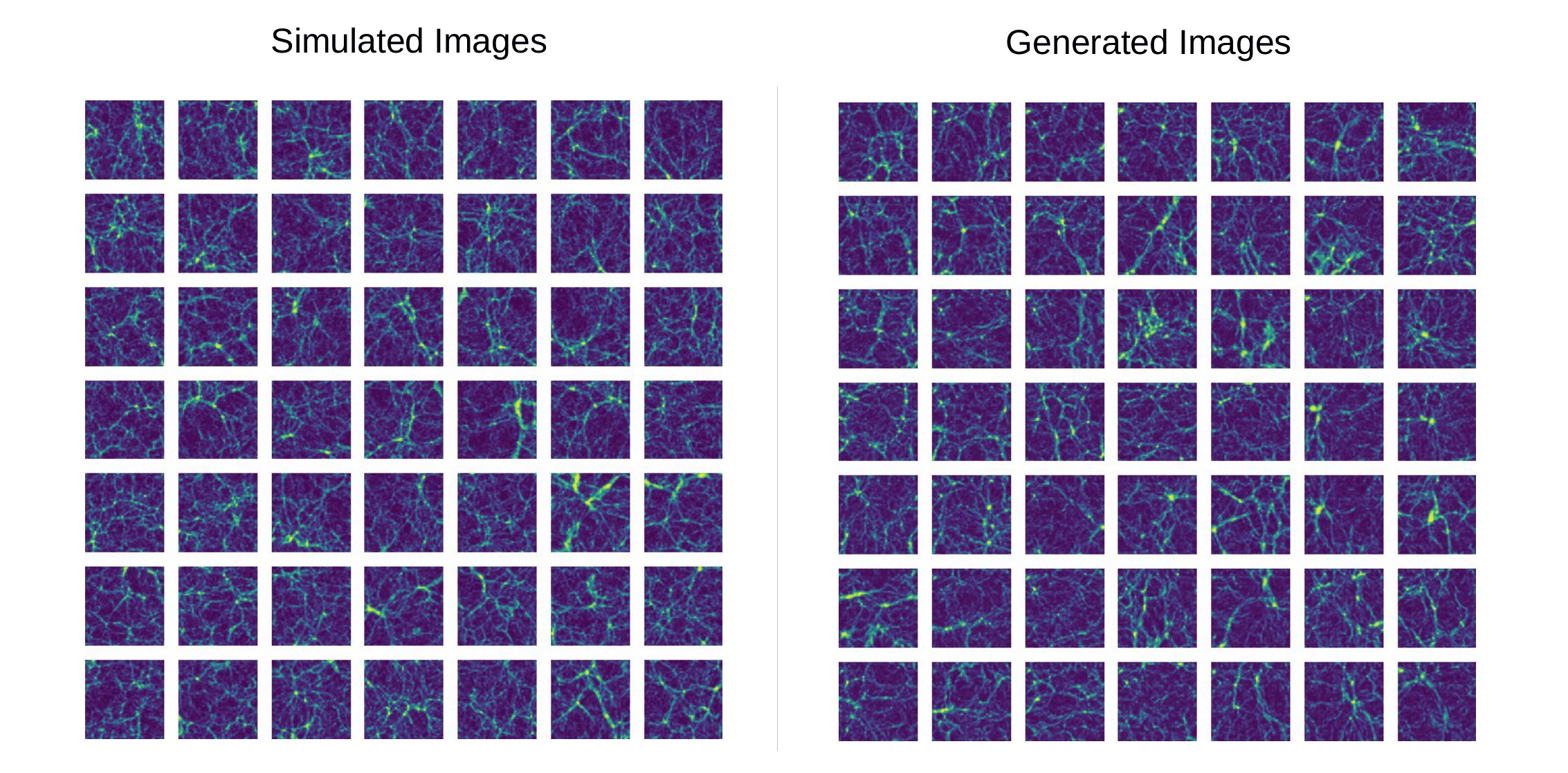} 
  \caption{Two subsets of 50 images taken at random from a set of 3D simulation images (left) and a set of images generated by the GAN (right). Every image represents a $128 \times 128$ log density map of side $50 Mpc$.}
  
  \label{fig:3D_GAN_ims}
\end{figure*}

As stated above, we need an additional test to quantify how well the autoencoder infers individual images. Therefore, we need a pairwise comparison between input images from the simulations and their inferred counterparts from the autoencoder.

Taking a simulation/inferred image pair, we test how well the structures overlap by thresholding the images at different values and counting the fraction of pixels above the threshold that overlap.

For a pair of images $a$ and $b$, the overlap is expressed in the following way:

\begin{equation}
O_{ab}(t)=\frac{N_{ab} (t)}{N_a(t)+ N_{b}(t)} \label{eq:simple_ovl}    
\end{equation}

Where $N_{a or b}(t)$ is the number of pixels whose value is above the threshold $t$ in $a$ or $b$ and $N_{ab}(t)$ is the number of pixels whose value is above $t$ for both $a$ and $b$ in a given position in an image. 

To get a sense of the overall quality of the encoded images, we plot the dice coefficient averaged over a set of simulated-inferred pairs: $\bar{O}(t)=\langle O_{ab}(t)\rangle_{a,b}$.

For clarity the overlap $O_{ab}(t)$ is plotted against the top percentage, associated to a given threshold, rather than the threshold itself (Fig. \ref{fig:2D_AE_overlap} and Fig. \ref{fig:3D_AE_overlap}). The thresholds are defined for the simulated and the inferred sets independently, i.e the top n\% of image a and b might correspond to slightly different thresholds.

It must be noted that a random pair of images will on average provide a non-zero overlap. Indeed, two images with $n\%$ thresholded pixels are expected to have an average overlap of $n\%$. To get a better sense of the entire inferred set's performance/recovery, we thus compute a \textit{random} overlap, as defined by an overlap measured over a random set of simulation pairs, and plot it along with our overlap averaged on simulation/inferred pairs.







From this random overlap measure we can proceed to build an unbiased estimator of feature recovery by subtracting it from the overlap measured for simulated/inferred image pairs.

By itself, this difference is not informative, therefore we must look at it relative to relevant values.

First we observe it relative to its maximum possible score $(1-r(t)$, $r(t)$ being the average random overlap for a given threshold $t$; this provides a completion score between 0 and 1, with 1 corresponding to a perfect overlap and 0 corresponding to a completely random overlap:

\begin{equation}
\bar{O}(t)_1=\frac{\bar{O}(t)-r(t)}{1-r(t)} \label{eq:score_ovl}
\end{equation}

We will refer to it as the normalized sd coefficient

Next we observe it relative to the standard deviation of the random overlap, to ensure whether or not the predictions, if imperfect, are well without the random range. This corresponds to a signal to noise ratio.

\begin{equation}
\bar{O}(t)_2=\frac{\bar{O}(t)-r(t)}{\sigma(r(t))} \label{eq:stn_ovl}
\end{equation}

We will refer to it as the sd coefficient significance.


These measures will allow us to determine whether the structures are well recovered, and at which scale they are best recovered.

\subsection{Results from GANs}\label{GANresults}

\begin{figure*}
  \centering
  \includegraphics[width=\textwidth]{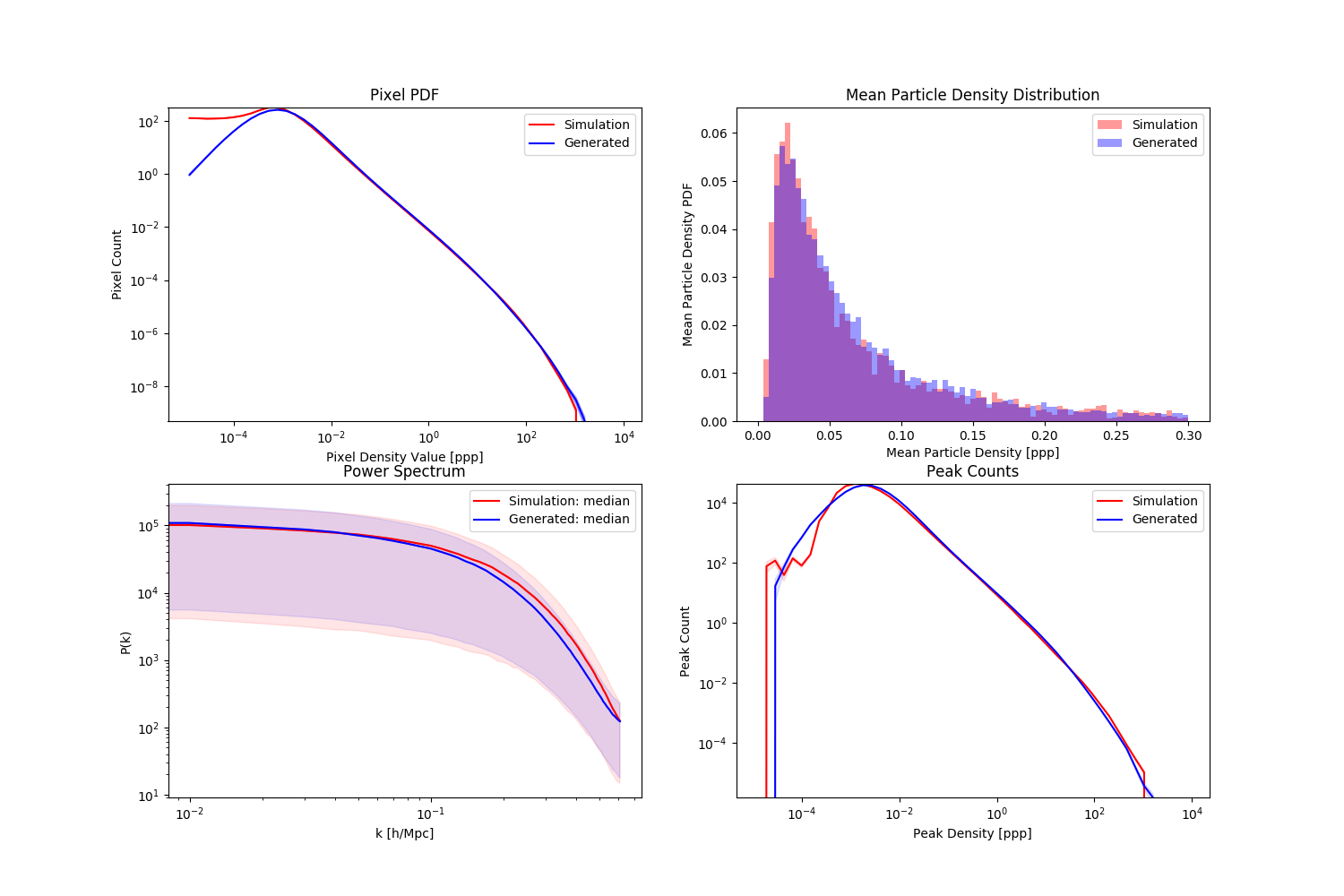} 
  \caption{Statistics of the 3D simulation images (red) compared to their GAN-generated counterparts (blue).Upper left shows the pixel PDF, upper right shows mean density distribution, lower left shows median power spectrum as well as mad (median  absolute deviation) layer, and lower right shows average peak count per image.}
  \label{fig:3D_GAN_stats}
\end{figure*}



\subsubsection{2D images}\label{GAN2D}

We first present the results of the GAN trained on images from 2D simulations. In our study, the GAN is trained for 60 epochs for best results but consistently outputs sets of very similar images as early as 20 epochs.


Two sets of 50 images taken at random from the simulations and from the GAN's generated images show, in Fig. \ref{fig:2D_GAN_ims}, the GAN's ability to generate images of convincing similarity. Visually, we  observe that the large scale structure is well recovered; this is most notably the case for the filaments, reproduced in all their diversity of length, thickness and frequency. It is also the case for high-density regions, or halos, in terms of their occurrence, brightness (or density) and positions within the structures. 

This observation is further corroborated by the statistical estimators as seen in Fig. \ref{fig:2D_GAN_stats}. The pixel PDFs (lower left panel of Fig. \ref{fig:2D_GAN_ims}) show a near-perfect overlap for the majority of the pixel density values, with a slight under-representation in the generated images of the densest values. This agreement shows that the density distribution of the images is very well recovered by the GAN.


The mean particle density distribution of the generated set (displayed in the upper right panel of Fig. \ref{fig:2D_GAN_ims}) shows a satisfactory agreement with the simulation set but exhibits a shift toward lower mean density values. The overall agreement indicates that the diversity of the original set is globally well represented in the generated set. The median power spectra and their \textit{mad} layer (Fig. \ref{fig:2D_GAN_stats} upper left panel), for both simulated and generated sets, show a satisfying overlap, indicating a good recovery of the correlations at various distances and thus a good representation of the different scales in the images. However, we note a slight downward shift for the generated images, which is due to a statistically lower overall density in these images, as could be seen in the mean particle density distribution.
Finally, in the lower right panel of Fig. \ref{fig:2D_GAN_ims}, we show the peak counts. The very good agreement between the true simulated images and the generated ones confirms that the dense halos are well represented. In particular, we can see that not only a similar average amount of peaks is observed but also a similar average distribution of the peak values on both sets, with a slight under-representation of the densest peaks and a more notable under-representation of low-density peaks. 
However the peaks at low density are due to image processing and not physical, so this is not an issue.

\subsubsection{3D images}\label{GAN3D}

We now turn to the results obtained by the GAN trained on the 3D simulated images. 
Once more the GAN progresses in a stable manner, consistently producing verisimilar images after around 20 epochs of training. For best results we train it for 70 epochs.




First we focus on two subsets taken at random from both the original set of simulated images and the set of generated images (Fig. \ref{fig:3D_GAN_ims}). Once again our visual inspection shows that the diversity of the simulated images is well recovered by the GAN in terms of distribution in size and frequency of filaments and number and brightness of high-density regions. A closer look at the statistical properties of the images as seen in Fig. \ref{fig:3D_GAN_stats} further confirms this.

Notably, the pixel PDFs (Fig. \ref{fig:3D_GAN_stats} upper left panel) show a near-perfect overlap, confirming the good recovery of the density distribution on the average images. However, the lower tail of the distribution is poorly represented for pixel values $<10^{-3}ppp$. The generated images show a deficit, while the simulations show a plateau. This can be explained by the saturation effect related to the constant $c$ in Eq. \ref{eq:rodriguez_tf}.







Meanwhile, the mean density PDF (Fig. \ref{fig:3D_GAN_stats} upper right panel) seems to be very well recovered, confirming the good recovery of the image diversity. The median power spectra and their \textit{mad} regions (Fig. \ref{fig:3D_GAN_stats} lower left panel) yield a near-perfect overlap, with a slight under-representation of higher frequencies in the generated images. In the lower right panel of Fig. \ref{fig:3D_GAN_stats}, we plot the peak counts. True simulated images and generated ones show once more near-perfect agreement, confirming that high-density region centers are well represented in terms of their numbers as well as their distribution. However, we observe a misrepresentation of the distribution lower tail similarly to the pixel PDF, for similar reasons.

For all of the estimators presented here and that are common with \cite{rodriguez2018fast} and \cite{feder2020nonlinear}, our findings agree with theirs. The mean density distributions of both simulated and generated sets are somewhat distinguishable but show a very good overlap, and the power spectra show a very satisfactory overlap between both their medians and their \textit{mad} regions. We note that our results seem coherent with those of \cite{feder2020nonlinear} who encountered similar saturation issues to ours at low densities for similar reasons. We also note that \cite{rodriguez2018fast} do not refer to a poor representation of tails in the pixel PDF and peak counts for their GAN-generated images, despite using a pixel transformation with a saturation effect.






\subsection{Results from the Autoencoder}\label{AEresults}

\subsubsection{2D images}\label{AE2D}

\begin{figure*}
  \centering
  \includegraphics[width=.8\textwidth]{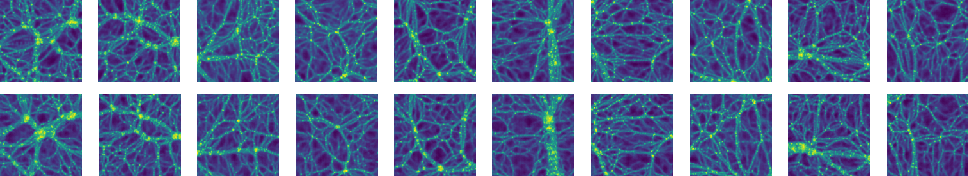} 
  \caption{10 images taken at random from the 2D simulations (top row) and their autoencoder-inferred counterparts (bottom row). Every image represents a $128\times128$ log density map of side $50 Mpc$.}
  \label{fig:2D_AE_ims}
  
\end{figure*}

\begin{figure*}
  \centering
  \includegraphics[width=\textwidth]{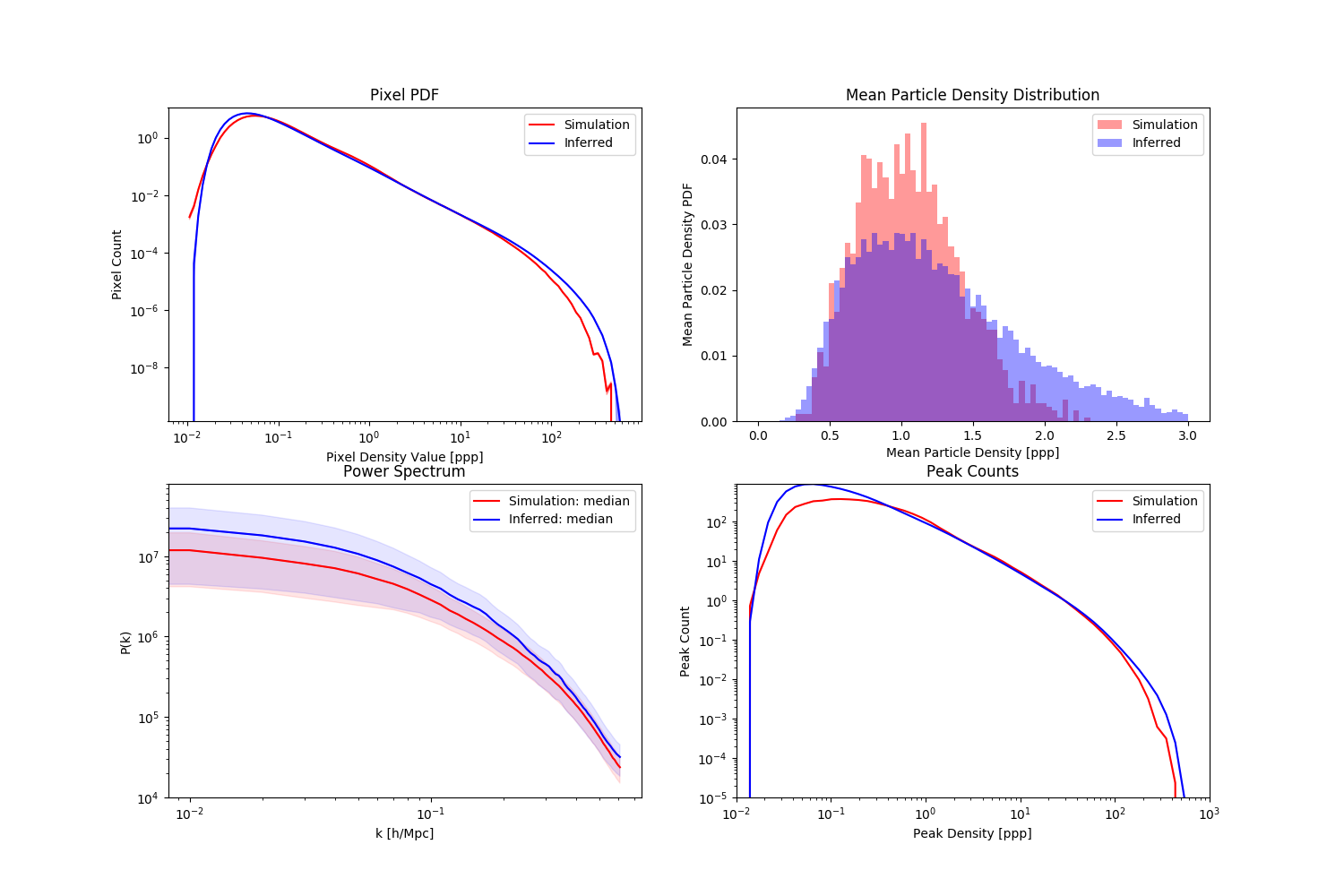} 
  \caption{Statistics of the 2D simulation images compared to their encoded counterparts. Upper left shows the pixel PDF, upper right shows mean density distribution, lower left shows median power spectrum as well as mad (median  absolute deviation) layer, and lower right shows average peak count per image.}
  
  \label{fig:2D_AE_stats}
  
\end{figure*}


We now focus on the outcome of the AE for the set of 2D images. For the 2D images, the autoencoder is trained over 50 epochs. We used the the loss function measured in the training set in order to avoid any overfitting, i.e. the point a network can reach during training at which it becomes too fine-tuned to its training set and starts to perform poorly on new sets. In practice, we observe the decrease of the loss function during the training process, oftentimes the loss measured on a separate \textit{validation} set reaches a point during training at which it starts to increase, which indicates overfitting. During our training of the AE on the 2D images we never observed an overfitting, the loss on the validation set near-monotonously converging  instead towards a constant.



We are interested in seeing how the AE fares with images it has never encountered during its training, as our goal is to be able to apply it on new datasets. As such all the images shown and used to measure the different statistical properties in the results are part of, or inferred from, a separate set than the ones used for training, called a {\it test set}. This will be the case for both 2D and 3D images.

%


\begin{figure*}
  \centering
  \includegraphics[width=.9\textwidth]{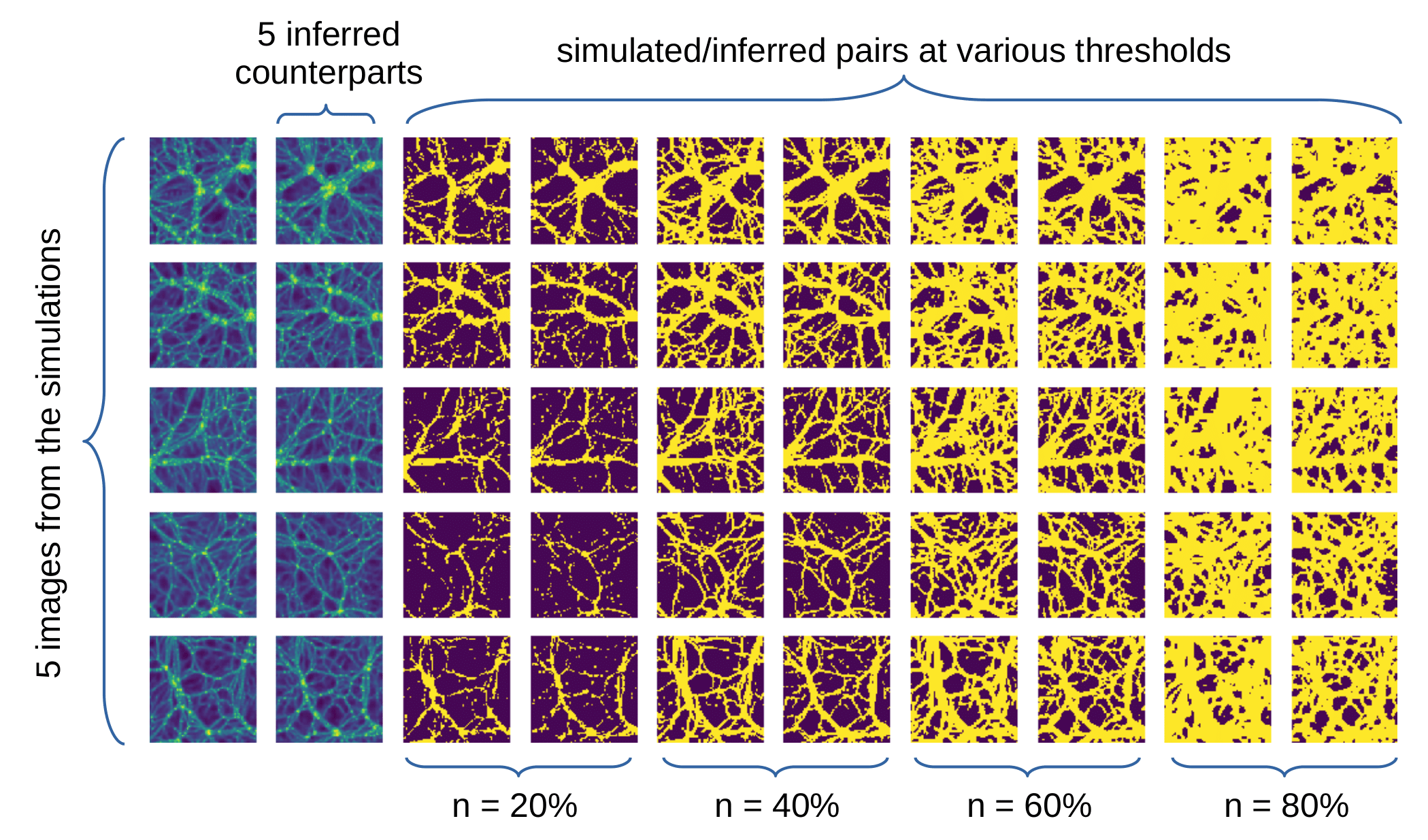} 
  \caption{Examples of 2D simulation/inferred image pairs (columns 1-2) and their thresholded equivalents as used in the computation of the overlap function. 
  The first column in each pair represents images from the simulations and the second their inferred counterparts.
  Here they are thresholded for top $20\%$ (c. 3-4), $40\%$ (c. 5-6), $60\%$ (c. 7-8), and $80\%$ (c. 9-10).}
  \label{fig:threshex}
  
\end{figure*}

We first illustrate the results in terms of the AE's performance and recovery of features with a set of ten simulated images taken at random from the test set (Fig. \ref{fig:2D_AE_ims}, first column from left to right) and their inferred counterparts (Fig. \ref{fig:2D_AE_ims}, second column from left to right). We note that the inferred images visually look similar to the simulated images but the larger and denser structures tend to be recovered better than the smaller diffuse structures.

We recall that while the decoder, having the exact same structure and weights as the GAN's generator, is expected to infer images that are statistically similar to the GAN's, it nevertheless infers images from a different prior. Indeed, while the GAN's inputs are selected randomly from a Gaussian distribution, the decoder's inputs are all constructed in a deliberate fashion by the AE's encoder and are not expected to follow the same distribution. As such, a change in statistics is not unexpected. 

As a sanity check, we additionally trained an isolated encoder directly on images randomly generated by the GAN, constraining it to output an encoded vector $z=E(G(x))$ similar to the generated images' input vector $x$, with a simple $l_{2}$ loss $L_{E}=\|z-x\|^2$. The ensuing images inferred by the AE when tested on new GAN-generated images give very satisfactory results with all estimators (near perfect overlap of all estimators, dice coefficient>0.7 even for the top 1\% pixels). However this did not translate well when testing the AE thus trained on real simulation images. This test suggests much of the AE's limitations might be due to a certain dissimilarity between GAN-generated images and true images, and thus due to the GAN's limitations themselves.


A closer inspection of the statistical properties of the images (Fig. \ref{fig:2D_AE_stats}) shows notable differences between the sets of inferred and simulated images. The pixel PDFs (Fig. \ref{fig:2D_AE_stats} upper left panel) show satisfactory overlap for the two sets but present a significant over-representation of high-density pixels in the inferred images. This over-representation of denser pixels becomes more noticeable in the mean particle density distribution (Fig. \ref{fig:2D_AE_stats} upper right panel), which exhibits a significant skewness towards higher mean densities. On the lower left panel of Fig. \ref{fig:2D_AE_stats}, the inferred images' median power spectrum shows a significant upward shift despite a similar shape, suggesting a similar distribution of frequencies but confirming once again the over-representation of higher density pixels. Finally and as expected, the peak counts (Fig. \ref{fig:2D_AE_stats} lower right panel) exhibit the same trend as the pixel PDF in terms of distribution of values. Thus, despite the seemingly good retrieval of the densest structures from visual inspection, the statistical properties suggest that the densest regions are in fact inferred with a higher density than in the input simulation images. 





\begin{figure}
  \centering
  \includegraphics[width=0.4\textwidth]{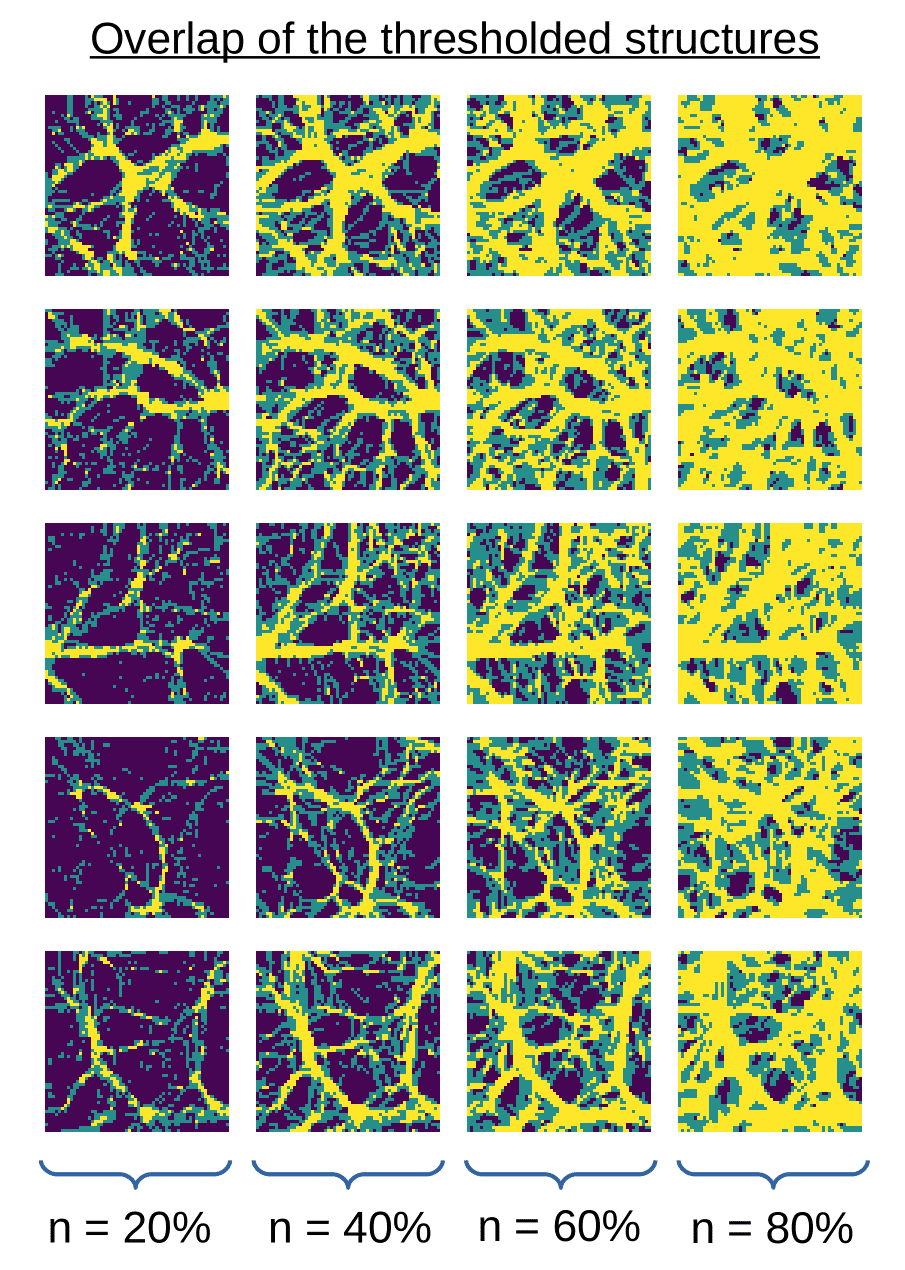} 
  \caption{Examples of the overlap of thresholded structures for 2D simulation/inferred image pairs. Yellow pixels indicate where the structures overlap and green where they do not. The dark background represents pixels below the threshold. The dice coefficient is simply measured as $\frac{n_{yellow}}{n_{yellow}+n_{green}}$ 
  Here they are thresholded for top $20\%$ (c. 1), $40\%$ (c. 2), $60\%$ (c. 3), and $80\%$ (c. 4).}
  \label{fig:overlapex}
\end{figure}

\begin{figure*}
  \centering
  \includegraphics[width=.8\textwidth]{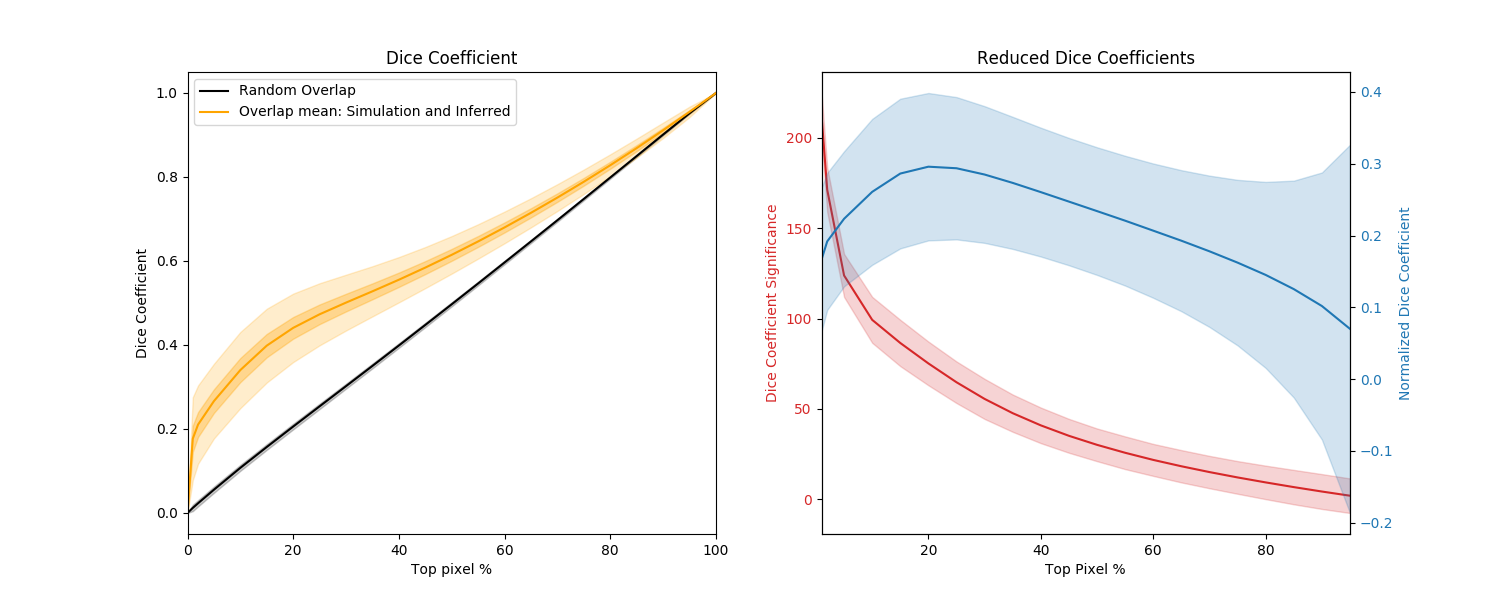} 
  \caption{Left: dice coefficient (as defined by Eq. \ref{eq:simple_ovl}) of top $n\%$ pixels between 2D simulation images and their inferred counterparts, by increments of $5\%$ (yellow). The inner dark yellow layer represents measure uncertainty and the outer yellow layer represents standard deviation over the set. Random overlap is represented in black. Right, red: dice coefficient significance (see Eq. \ref{eq:stn_ovl}), blue: normalized dice coefficient (see Eq. \ref{eq:score_ovl}); both are represented with their standard deviation layers.}
  \label{fig:2D_AE_overlap}
  
\end{figure*}


We now focus on the Sørensen-Dice coefficient which computes the overlap fraction of two thresholded images. 
Visual inspection (Fig. \ref{fig:threshex} and Fig. \ref{fig:overlapex}) of the thresholded simulated/inferred image pairs and how they overlap suggests that dense structures are strikingly well recovered, with the autoencoder favoring the retrieval of thick, contrasted features to the detriment of finer ones. Once again, this is expected given the CNN's predisposition to detect and construct well-defined shapes.


An inspection of the dice coefficient  (Fig. \ref{fig:2D_AE_overlap}, left) corroborates this finding . Indeed, despite some slight shifts of structures and aforementioned loss of finer structures, we note that the high density regions overlap satisfyingly and seemingly well without the random region for up to the top $60\%$ pixels. The normalized dice coefficient (Fig. \ref{fig:2D_AE_overlap}, right, blue) lets us assess the density threshold at which the autoencoder best captures structures. Here, it peaks for the top $20\%$ pixels.

\subsubsection{3D images}\label{AE3D}

\begin{figure*}
  \centering
  \includegraphics[width=\textwidth]{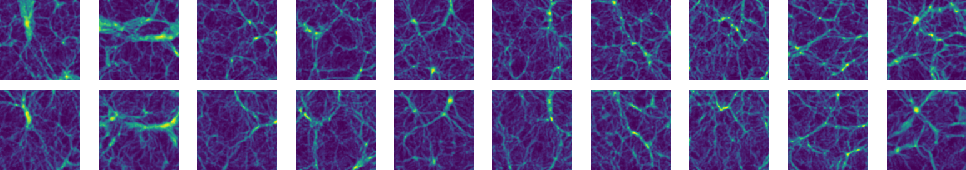} 
  \caption{10 images from the 3D simulations (top row) and their 10 autoencoder-inferred counterparts (bottom row). Every image represents a $128\times128$ log density map of side $50 Mpc$.}
  \label{fig:3D_AE_ims}
  
\end{figure*}

\begin{figure*}
  \centering
  \includegraphics[width=\textwidth]{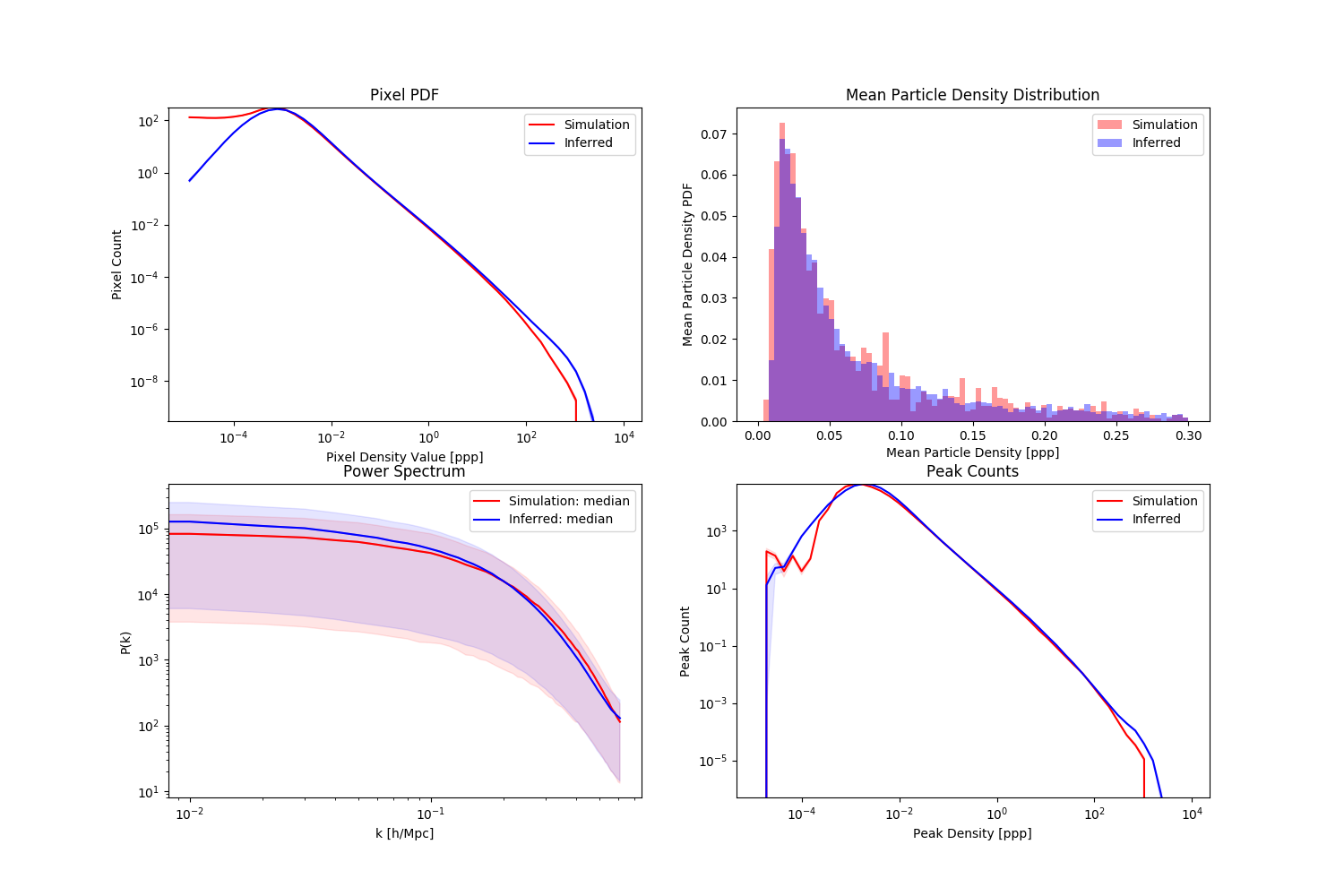} 
  \caption{Statistics for the 3D simulation images (red) and their AE-inferred counterparts (blue). Upper left shows the pixel PDF, upper right shows mean density distribution, lower left shows median power spectrum as well as mad (median  absolute deviation) layer, and lower right shows average peak count per image. }
  \label{fig:3D_AE_stats}
  
\end{figure*}

We now turn to the AE's performance when trained on images from the 3D simulations. For these images, the autoencoder is trained over 50 epochs but the validation loss shows that overfitting starts slowly at epoch 25. Therefore, we keep the network's weights saved at that time and analyse the results for this set of weights.

We first select at random ten images from the 3D simulations test-set and their inferred counterparts (two left-most columns of Fig. \ref{fig:3D_AE_ims}). Once again, we can observe that the densest structures seem to be better recovered than the more diffuse ones. We concentrate on the images' statistical properties (Fig. \ref{fig:3D_AE_stats}) to better assess the AE's performance. 

First, we see that the pixel PDFs (Fig. \ref{fig:3D_AE_stats} upper left panel), as for the GAN case, overlap well up to the lower tail of the distribution, with an under-representation of lower densities. However in this case, the higher densities are slightly over-represented in the inferred images. The mean density distribution, as can be seen in the upper right panel of Fig. \ref{fig:3D_AE_stats}, is well recovered. The over-representation in higher densities causes a slight upward shift of the inferred images' power spectra (Fig. \ref{fig:3D_AE_stats} lower left). As for the peak counts (Fig. \ref{fig:3D_AE_stats} lower right panel), they show similarly to the pixel PDF a slight over-representation in the high-density regime.

\begin{figure*}
  \centering
  \includegraphics[width=.8\textwidth]{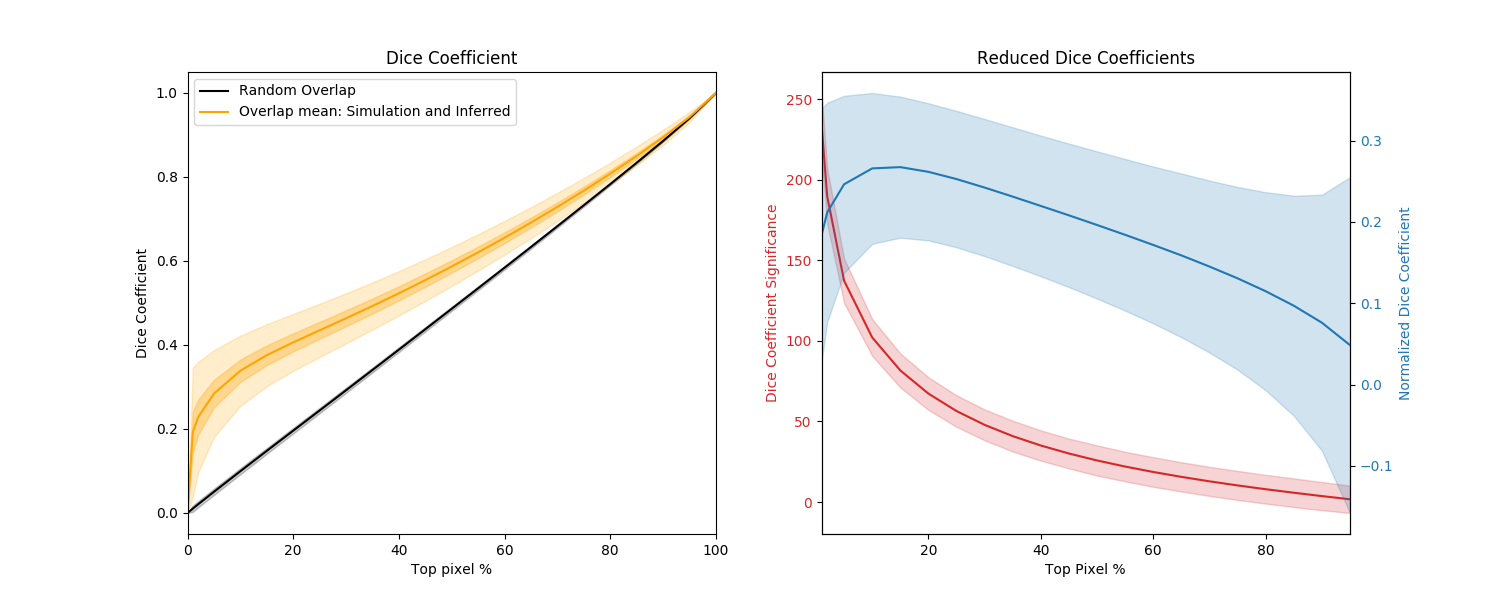} 
  \caption{Left: dice coefficient (as defined by Eq. \ref{eq:simple_ovl}) of top $n\%$ pixels between 2D simulation images and their inferred counterparts, by increments of $5\%$ (yellow). The inner dark yellow layer represents measure uncertainty and the outer yellow layer represents standard deviation over the set. Random overlap is represented in black. Right, red: dice coefficient significance (see Eq. \ref{eq:stn_ovl}), blue: normalized dice coefficient (see Eq. \ref{eq:score_ovl}); both are represented with their standard deviation layers.}
  \label{fig:3D_AE_overlap}
  
\end{figure*}

An inspection of the dice coefficient  (Fig. \ref{fig:3D_AE_overlap}, left) shows an overall satisfying recovery of the images which significantly differ from random images. We further note that the retrieval of high-density structures is good for the top $60\%$ pixels in a majority of images as exhibited by the dice coefficient significance (Fig. \ref{fig:3D_AE_overlap}, right, red). The normalized dice coefficient (Fig. \ref{fig:3D_AE_overlap}, right, blue) suggests that structures associated with the top $10\%$ pixels are the ones that are best captured by the autoencoder.

\section{Conclusion} \label{conclusion}

We trained a GAN on images derived from two types of cosmic web simulations, to produce statistically consistent images. The first set was built from 2D N-body simulations using a field estimation method described in \cite{aragoncalvo2020smooth}, and the second set from 3D simulations using a smoothed-histogram field estimation method.

Using an ensemble of estimators (pixel PDF, mean density distribution, power spectrum and peak counts), we confirmed the GAN's ability to extract the underlying statistical distribution of images built from the simulations and generate new images hailing from this distribution. We showed that this was the case for both image types. They were indeed emulated with striking similarity to the true original ones as was shown both visually and via the near-perfect overlap of the different statistical estimators. Additionally, the training proved stable with the networks consistently generating images of increasing quality with training up to a stable point where the generated images are visually indistinguishable from the true ones.

We note that despite the success of GANs to reproduce with high fidelity a desired input, it is important to be careful when using these black box models. As we noted when looking at some of the statistical estimators such as the pixel PDF, a visual inspection is not sufficient to guarantee that the generated dataset is statistically equivalent to the targeted one. Since the generated images cannot be distinguished visually from the true ones, it is thus very important to design a series of quantitative tests ensuring in which range they can be exploited. In other words, neural networks are very appealing tools to be explored and exploited but at the cost of devising precise tests of their domains of validity and their generalization capability.

Building on the GAN's properties and as a first step towards constructing a predictive model, we used the trained network to devise an autoencoder that reduces $128 \times 128$ pixel images, and related information, to encoded vectors of smaller size (100 elements in our case) which can then be decoded with as little loss as possible. 

A visual appraisal of the images inferred by the autoencoder suggested that the large dense structures were very well reproduced for both image types (2D- and 3D-derived), while the smaller fine structures were reconstructed more randomly.
Furthermore, the inferred images seemed visually realistic; however statistical estimators suggested an overall decrease in statistical similarity with the original sets, when compared with the GAN.
This is to be expected; although the decoder is constrained to output realistic images when given a Gaussian-distributed input, here the inputs supplied by the encoder did not follow such a distribution.

Finally, the fact that the autoencoder manages to reproduce successfully the contrasted features of the input images indicates that the GAN did capture meaningful features of the dataset, and is far from overfitting it by only reproducing the learning dataset with small perturbations.


Given our autoencoder's ability to extract an image's essential information and reproduce the image based on this information, it opens the way to the exploitation of these properties for future work. For instance, GANs have been used to do inpainting or denoising \citep{cheng2019misgan}. In our setting, it would be possible to take advantage of the learned features on a simulated dataset to deal with noise and missing values in observational data. The autoencoders could also be used for predictive purposes, such as inferring $z=0$ from simulation images earlier in time (higher $z$). In this last case, observation at a given $z$ could therefore serve to predict the future or the past for our universe.


\begin{acknowledgements}
      The authors thank M. Aragon-Calvo for providing 2D simulations, H. Tanimura for providing the 3D simulations, and the whole ByoPiC team for useful discussions. This project was funded by the European Research Council (ERC) under the European Union's Horizon 2020 research and innovation programme grant agreement ERC-2015-AdG 695561. A.D. was supported by the Comunidad de Madrid and the Complutense University of Madrid (Spain) through the Atracción de Talento program (Ref. 2019-T1/TIC-13298).
\end{acknowledgements}

\bibliographystyle{aa}
\bibliography{references} 

\begin{appendix}

\section{Network Specifications}\label{sec:appenA}

\subsection{GAN}

The Generator and Discriminator are built following the structure detailed in \ref{table:archiGAN}.
They are trained using the {\it Adam} optimizer with parameters ($lr=0.0002, \beta_1=0.5$) and minimize the loss given in eq. \ref{eq:logloss} wherein a small noisy component that gives the images the wrong label is added. This imperfect loss avoids the common pitfall of GANs wherein one of the competing parties, usually the discriminator, becomes too efficient compared to the other, stopping the competition and thus the training well before quality images can be generated.

\subsection{AE}

We remind that Autoencoder is built by appending a second network, the Decoder, to a first, the Encoder. In our case as explained in \ref{AE}, we take a GAN's trained generator as a readily built decoder. Similarly, we base the encoder's architecture on the discriminator's, changing only the final $(8192 \longrightarrow 1)$ dense layer to a $(8192 \longrightarrow 100)$ dense layer with no activation, such that the output is of the correct size and "limitations". As such the general structures of both the encoder and decoder are also defined in \ref{table:archiGAN}. With the weights of the decoder fixed, the AE is trained by varying the weights of the encoder to reduce the loss given in \ref{eq:TDloss} using the {\it Adam} optimizer with the same parameters as above ($lr=0.0002, \beta_1=0.5$).

\begin{figure}[h!]\label{table:archiGAN}
    \centering
\begin{tabular}{|c|c|} 
 \hline
 Filter sizes & $\{5, 5, 5, 5, 5\}$ \\
 \hline
 $n_{filter} (G/De)$  & $\{256, 128, 64, 32, 1\}$ \\
 \hline
 $n_{filter} (D/E)$ & $\{32, 64, 128, 256, 512\}$ \\
 \hline
 Strides:  &  $\{2, 2, 2, 2, 2\} $ \\
 \hline
 Padding:  &  $ \{0, 0, 0, 0, 1\}  *$\\
 \hline
 Layer Act. & ReLU (G/De), Leaky ReLU (D/E) \\
 \hline
 Final Act. & Tanh (G/De), Sigmoid (D), None(E) \\
 \hline
 Latent dimension & $100$ \\
 \hline
\end{tabular}

    \caption{Architecture specifications for each layer of the GAN's generator network (G) and discriminator network (D) as well as the AE's encoder (E) and decoder(De). Asterisk signifies that order of layers is reversed in the discriminator and Encoder.}
    \label{fig:my_label}
\end{figure}

\end{appendix}

\end{document}